\documentclass{ws-procs975x65}

\newcommand{\be}{\begin{equation}}
\newcommand{\ee}{\end{equation}}
\newcommand{\bea}{\begin{eqnarray}}
\newcommand{\eea}{\end{eqnarray}}
\newcommand{\I}{\rm i}
\newcommand{\lb}{\label}

\usepackage{color}

\begin{document}

\title{Quantum Cosmology for the XXI$^{\mathrm{st}}$ Century: A Debate}

\author{Martin Bojowald$^1$}
\address{The Pennsylvania State University,
Institute for Gravitation and the Cosmos\\
University Park, PA 16802, USA}

\vspace*{-0.3cm}\author{Claus Kiefer$^2$}
\address{Institut f\"ur Theoretische Physik, Universit\"{a}t zu
K\"{o}ln, \\ Z\"{u}lpicher Strasse 77, 50937 K\"{o}ln, Germany}

\vspace*{-0.3cm}\author{Paulo Vargas Moniz$^3$}

\address{Departamento de F\'{\i}sica, Universidade da Beira
Interior, \\ Rua Marqu\^{e}s d'Avila e Bolama, 6200 Covilh\~{a},
Portugal\\\vspace*{0.5cm}E-mails:
\mbox{$^1$bojowald@gravity.psu.edu},
\mbox{$^2$kiefer@thp.uni-koeln.de},
\mbox{$^3$pmoniz@ubi.pt}}

\begin{abstract}
  Quantum cosmology  from the late sixties into the early
  XXI$^{st}$ century is reviewed and  appraised
    in the form
  of a debate, set up by two presentations on mainly the Wheeler--DeWitt
  quantization and on loop quantum cosmology, respectively.  (Open)
  questions,
    encouragement
    and guiding lines
    shared
  with the audience
    are provided here.
  \end{abstract}


\bodymatter

\section{\emph{Raison d'\^{e}tre} (Paulo Vargas Moniz)} \label{sec1}

\indent

Quantum cosmology is the application of quantum theory to the Universe
as a whole. As such, it plays a special role in physics. First, it is
based on a quantum theory of gravity about which no general agreement
has been reached. Second, as part of cosmology, it has not yet been
susceptible to observational tests.

Providing a corresponding summary was a major aim of the debate
reported here\footnote{
I
decided on a more pro-active and all-inclusive dynamical discussion
of quantum cosmology.
  Quite like the ``Presidential debates''
that were seen
  in the run for office at the White House\cite{wiki}, in which an
  audience of citizens could directly
  put questions.
  }. More precisely, we wanted to present a viewpoint on
past progress, the current situation, and new approaches to
proceed\cite{OUP}. Hence the presentation by Claus Kiefer as
summarized in Sec.~\ref{sec2}. On the other hand, we intended this
to be an interchange of ideas, dynamically including and challenging
the audience.

A subsequent talk was delivered by Martin Bojowald (see
Sec.~\ref{sec3}), in which the more recent progress of a new
school\cite{LivRev}
--- bringing novel ideas into the current century, together with
methods as well as challenges to quantum cosmology --- was widely
acknowledged: by importing techniques and results from loop quantum
gravity, quantum cosmology did gain new momentum, and new results for
quantum cosmology have been appearing in the literature ever since.

The subject of quantum cosmology therefore enjoys renewed
interest, but does perhaps not involve as large a set of groups and
researchers as it deserves. For this reason the debate, whose two lecturers
are not in opposing camps, but rather act as colleagues striving for
progress within overlapping schools, has left ample time
for open questions and fair criticism as
well as some prudent scepticism from the audience. The overarching
question of the debate was: can quantum cosmology take a main stance and
central stage in XXI$^{\mathrm{st}}$-century research? A
(tentative!) summary of the questions and constructive replies is
collected in Sec.~\ref{sec4}.

In a slowly progressing, only thinly populated research area,
earlier results are often forgotten, overlooked or ignored, then
rediscovered, rederived or reproduced in disguise. An appraisal of
preceding contributions may, at a pragmatic level, save some work
and, more idealistically, provide unity. This is one of our aims.
We invite all
unhappy readers of these notes to send us comments about serious
omissions or misrepresentations, which may find a place in a future
review.

Was this debate worth it? We think so. We add some thoughts in
Sec.~\ref{sec5}, proposing where new directions could be sought,
even if other results from other sometimes and somewhat unforeseen
directions may turn up in the meantime, then further promote the
subject and continue to lead its progress. But anyway, that is how
research most of the time makes significant progress:
$\mathbf{Unexpectedly}$.

%%%%%%%%%%%%%%%%%%%%%%%%%%%%%%%%%%%%%%%%%%%%%%%%%%%%%%%%%%%%%%%%%%

\section{A legacy from the XX$^{\mathrm{th}}$ century (Claus Kiefer)}
\label{sec2}

\subsection{Why quantum cosmology?}

Why should or even must one apply quantum theory to the Universe as a whole?
Is it not sufficient to consider the standard picture of
cosmology describing the
Universe as expanding from a dense hot phase in the past to its
present state with galaxies and clusters of galaxies? After all, there
exist at present no observations for which a quantum cosmological
explanation seems to be mandatory.

There is, in fact, support for quantum cosmology for at least two
reasons.\footnote{In the following, I shall heavily rely on my earlier
  contributions to this topic, in particular on [\refcite{OUP}] as
  well as on [\refcite{KS08,vaas2,bremen}] and the references therein.
  A classic introduction to quantum cosmology is [\refcite{Halliwell}].}
First, general relativity is incomplete in that it predicts the
occurrence of singularities in a wide range of situations. This
concerns the origin of the Universe (``Big-Bang'') but also its
final fate; some models using dark energy as an explanation for the
current acceleration of the Universe predict singularities in the
future. Therefore, a more general theory is needed in order to
encompass these situations. The general belief is that this theory
is a quantum theory of gravity, for it was, after all, quantum
mechanics which rescued the atom from the singularities of classical
electrodynamics.

 The second reason derives from a general feature of quantum
theory. Except in microscopic cases, most quantum systems cannot be
considered as isolated. They interact with their natural
environment, as a result of which a globally entangled state results
that includes the variables of the system and the environment. For
macroscopic systems, this entanglement leads to the emergence of
classical properties for the system -- a process called {\em
decoherence} \cite{deco}. Since the environment of a system is again
coupled to its environment, the only truly closed quantum system is
the Universe as a whole. One arrives in this way straightforwardly
at the notion of a ``wave function of the Universe''. This notion is
conceptually independent of any particular interaction and a direct
consequence of the central feature of any quantum theory --
entanglement. As such it was already employed, at least in spirit,
by Hugh Everett in his presentation of the ``relative-state
interpretation'' \cite{Everett} and can be found also at other
places.\footnote{To quote from the monograph\cite{FH65} of Feynman
and Hibbs, p.~58: ``All of history's effect upon the future of the
Universe could be obtained from a single gigantic wave function.''}

Since gravity is the dominating interaction at cosmic scales, quantum
cosmology must, however, be based on a theory of quantum gravity
\cite{OUP}. But which theory?

\subsection{Which framework should one use?}

At present we do not have a final framework for a quantum theory of
gravity. Among the existing approaches, one can
mainly distinguish between the direct quantization of Einstein's
theory of general relativity and string theory (or M-theory). The
latter is more ambitious in the sense that it aims at a unification
of all interactions within a single quantum framework.
Quantum general relativity, on the other hand, attempts to
construct a consistent, non-perturbative, quantum theory of the
gravitational field on its own.

The fundamental length scales that are connected with these theories
are the Planck length, $l_{\rm P}=\sqrt{G\hbar}$, or the string
length, $l_{\rm s}$.\footnote{We set $c=1$ throughout.} It is
generally assumed that the string length is somewhat larger than the
Planck length. Although not fully established in quantitative
detail, quantum general relativity should follow from superstring
theory for scales $l\gg l_{\rm s}>l_{\rm P}$. Can one, in spite of
this uncertainty about the fundamental theory, say something
reliable about quantum gravity?  In [\refcite{GRG}] I have made the
point that this is indeed possible. The situation can be compared to
the role of the quantum mechanical Schr\"odinger equation. Although
this equation is not fundamental (it is non-relativistic, it is not
field-theoretic), important insights can be drawn from it. For
example, in the case of the hydrogen atom, one has to impose
boundary conditions for the wave function at the origin $r\to 0$,
that is, at the centre of the atom. This is certainly not a region
where one would expect non-relativistic quantum mechanics to be
exactly valid, but its consequences, in particular the resulting
spectrum, are empirically correct to an excellent approximation.

Erwin Schr\"odinger has found his equation by ``guessing'' a wave equation
from which the Hamilton--Jacobi equation of classical mechanics can be
recovered in the limit of small wavelengths, analogously to the limit
of geometric optics from wave optics. The same approach can be applied
to general relativity. One can start from the Hamilton--Jacobi version
of Einstein's equations and ``guess'' a wave equation from which they
can be recovered in the classical limit. The only assumption that is
required is the universal validity of quantum theory, that is, its
linear structure. It is not yet needed for this step to impose a
Hilbert-space structure (a linear space with a scalar product).
 Such a structure is employed in quantum
mechanics because of the probability interpretation for which one
needs a scalar product and its conservation in time (unitarity). The
status of this interpretation in quantum gravity remains open. We
should, however, keep in mind that it is exactly the normalization of
quantum states which is crucial in obtaining the correct spectra for
atoms and other systems.

The result of this approach is quantum geometrodynamics. Its central
equation is the Wheeler--DeWitt equation, first discussed by Bryce
DeWitt and John Wheeler in the 1960s. In a shorthand notation, it is
of the form \be \label{WdW} {\mathcal{H}}\Psi=0\ , \ee where
${\mathcal{H}}$ denotes the full Hamiltonian for both the
gravitational field (here described by the three-metric) as well as
all non-gravitational fields.\footnote{Strictly speaking, one has
the quantized Hamiltonian
  constraint as well as quantized diffeomorphism
  constraints.}
 For the detailed structure of this equation I can refer, for
example, to the classic paper by DeWitt \cite{DeWitt} and Wheeler
\cite{Wheeler} or my review in [\refcite{OUP}]. Two properties are
especially important for our purpose here. First, this equation does
not contain any classical time parameter $t$. The reason is that
space-time as such has disappeared in the same way as particle
trajectories have disappeared in quantum mechanics; here, only space
(the three-geometry) remains. Second, inspection of ${\mathcal{H}}$
exhibits the local hyperbolic structure of the Hamiltonian, that is,
the Wheeler--DeWitt equation possesses locally the structure of a
Klein--Gordon equation (that is, a wave equation). In the vicinity
of Friedmann Universes, this hyperbolic structure is not only
locally present, but also globally. One can thus define a new time
variable which exists only intrinsically and which can be
constructed from the three-metric (and non-gravitational fields)
itself. It is this absence of external time that could render the
probability interpretation and the ensuing Hilbert-space structure
obsolete in quantum gravity, for no conservation of probability may
be needed.\footnote{The situation is different for an isolated
quantum
  gravitational system such as a black hole; there, the semiclassical
  time of the rest of the Universe enters the description
  \cite{KMM09}.}
Of course, a Hilbert-space structure {\em is} needed in the
semiclassical limit discussed below.

In the following I shall briefly review the key points in the
application of the Wheeler--DeWitt equation to quantum cosmology.

\subsection{Wheeler--DeWitt equation and boundary conditions}

Cosmology can only be dealt with if one makes simplifying
assumptions. Since the Universe looks approximately homogeneous and
isotropic on large scales, one can impose this assumption on the
metric of space-time. As a result, one obtains the
Friedmann--Lema\^{\i}tre models usually employed. Such models are
called ``minisuperspace models''.

In the case of a Friedmann Universe with a homogeneous scalar field
$\phi$, the Wheeler--DeWitt equation reads (see e.g.
[\refcite{OUP,Halliwell}] or the Appendix of [\refcite{KS08}] for a
derivation) \be \lb{whdw1} {\mathcal H}\Psi=\left\{ \frac{2\pi
    G\hbar^2}{3}\frac{\partial^2}{\partial\alpha^2}-
  \frac{\hbar^2}{2}\frac{\partial^2}{\partial\phi^2}
+ {\rm e}^{6\alpha}\left(V\left(\phi\right)+\frac{\Lambda}{8\pi
      G}\right) -3{\rm e}^{4\alpha}\frac{k}{8\pi
    G}\right\}\Psi(\alpha,\phi)=0\ , \ee with cosmological constant
$\Lambda$ and curvature index $k=\pm 1, 0$. The variable $\alpha=\ln
a$, where $a$ stands for the scale factor, is introduced to obtain a
convenient form of the equation.

The general structure of the Wheeler--DeWitt equation concerning the
concept of time produces a peculiar notion of determinism at the
level of quantum cosmology. Despite the absence of an {\it external}
time parameter, the equation is of hyperbolic form thus suggesting
to use the $3$-volume $v$ or $\alpha=\frac13\ln v$ as an {\it
intrinsic} time parameter. Exchanging the classical differential
equations in time for a differential equation hyperbolic in $\alpha$
alters the determinism of the theory: the wave function is evolved
from small $\alpha$ to large $\alpha$, but not along a classical
trajectory parametrized by $t$. This has important consequences for
a classically recollapsing Universe, because in the quantum theory
both Big-Bang and Big-Crunch correspond to $\alpha\to-\infty$ and
are thus conceptually indistinguishable.

Implementing boundary conditions in
quantum cosmology differs from the situation in both general
relativity and ordinary quantum mechanics. In the following I shall
briefly review two of the most widely discussed boundary conditions:
the ``no-boundary proposal'' and the ``tunnelling proposal'' \cite{OUP}.

Also called the ``Hartle--Hawking proposal'' \cite{HH}, the
no-boundary proposal is essentially of a topological nature. It is
originally based on a Euclidean path integral representation for the
wave function, in which it is assumed that the integration is over
compact manifolds with only one boundary -- the boundary on which
the wave functional is defined. The term ``no-boundary proposal''
arises from the fact that there is no other boundary; there is, in
particular, no boundary corresponding to $a\to 0$. In order to
guarantee convergence, it is in general necessary to integrate over
complex metrics and to associate distinguished solutions with
particular contours in the complex plane \cite{contours}.

Except for the simplest cases, the path integral cannot be evaluated
exactly. One therefore has to resort to semiclassical (``saddle
point'') approximations. For the above model (\ref{whdw1}), taking
$\Lambda=0$, one gets in this approximation the following expression
for the wave function (here, $\hbar=1=G$), \be \label{NB} \Psi_{\rm
NB}\propto \left(a^2V(\phi)-1\right)^{-1/4}
\exp\left(\frac{1}{3V(\phi)}
\right)\cos\left(\frac{(a^2V(\phi)-1)^{3/2}}{3V(\phi)}-\frac{\pi}{4}\right)\
. \ee

The tunnelling proposal emerged from the work by Alexander Vilenkin
and others, cf. [\refcite{vilenkin,BKKS}] and references therein. It
is most easily formulated in mini\-superspace. In analogy with, for
example, the process of $\alpha$-decay in quantum mechanics, it is
proposed that the wave function consists solely of {\em outgoing}
modes. ``Outgoing'' means that the sign of the phase in the
wave function is distinguished from the outset. In contrast to the
no-boundary proposal, the tunnelling proposal thus leads to complex
wave functions.

In the above model one get the following wave function: \be
\Psi_{\rm T}\propto
(a^2V(\phi)-1)^{-1/4}\exp\left(-\frac{1}{3V(\phi)}
\right)\exp\left(-\frac{\I}{3V(\phi)}(a^2V(\phi)-1)^{3/2}\right)\ .
\ee Consequences of this difference to (\ref{NB}) arise, for
example, if one asks for the probability of an inflationary phase to
occur in the early Universe: whereas the tunnelling proposal seems
to favour the occurrence of such a phase, the no-boundary proposal
seems to disfavour it. No final word on this issue has, however,
been spoken. It is interesting
 that the tunnelling proposal allows the possibility that the Standard-Model
 Higgs field can play the role of the inflaton if a nonminimal
 coupling of the Higgs field to gravity is invoked \cite{BKKS}.

The application of these boundary conditions (as well as others
discussed in the literature) is thus mainly restricted to the
semiclassical realm. What one would like to do is to formulate a
proper boundary condition for the Wheeler--DeWitt equation
(\ref{whdw1}), that is, formulate a boundary condition from which a
unique solution follows. This is by no means a simple task
\cite{OUP}. For the class of hyperbolic partial differential
equations, to which (\ref{whdw1}) belongs, a proper boundary value
problem is the Cauchy problem: specify $\psi$ and
$\partial\psi/\partial\alpha$ at constant scale factor, and a unique
solution results. This is, however, not a proper boundary problem in
the case of classically recollapsing Universes. There, the
wave function must tend to zero for large scale factor; otherwise,
the correspondence with the classical theory is lost. The
specification of $\psi$ at $a=$ constant and demanding $\psi\to 0$
for $a\to\infty$ will, however, not lead to a unique solution. It
will in general not lead to any solution at all unless specific
(``quantized'') values for the parameters are chosen, see, for
example, the simple model discussed in [\refcite{CK90}]. A
mathematical discussion of these issues can be found in
[\refcite{Gerhardt}].

\subsection{Inclusion of inhomogeneities and the semiclassical
  picture}

Realistic models require the inclusion of further degrees of freedom;
after all, our Universe is not homogeneous. This is usually done by adding
a large number of multipoles describing density perturbations and
small gravitational waves \cite{HH85,OUP}. One can then derive an
approximate Schr\"odinger equation for these multipoles, in which the
time parameter $t$ is defined through the mini\-superspace variables
(here, $a$ and $\phi$). The derivation is performed by a
Born--Oppenheimer type of approximation scheme.
The result is that the total state (a solution of the Wheeler--DeWitt
equation) is of the form
\begin{equation}
\label{expiS}
\Psi \approx C(a,\phi)\exp({\rm i}S_0(a,\phi)/\hbar) \, \prod_n\psi_n(a,\phi,x_n)\ ,
\end{equation}
where $\{ x_n\}$ stands for the inhomogeneities (``multipoles'').
 In short, one has that
\begin{itemize}
\item $S_0$ obeys the Hamilton--Jacobi equation for $a$ and $\phi$
and thereby defines a classical space-time which is a solution to
Einstein's equations (this order is formally similar to the recovery
of geometrical optics from wave optics via the eikonal equation);
$C(a,\phi)$ denotes a slowly varying amplitude.
\item The multipole wave functions $\psi_n$ obey approximate
  Schr\"odinger equations,
\begin{equation}
\label{semi}
 {\rm i}\hbar \frac{\displaystyle\partial\psi_n}{\displaystyle\partial
t}:= {\rm i}\hbar \, \nabla \, S_0 \, \cdot\, \nabla
\psi_n \approx {\mathcal{H}}_n \, \psi_n \ ,
\end{equation}
where the ${\mathcal{H}}_n$
denote the Hamiltonians for the multipole degrees of freedom. The
$\nabla$-operator on the left-hand side of (\ref{semi}) is a shorthand
notation for derivatives with respect to the minisuperspace variables
(here: $a$ and $\phi$).  Semiclassical time $t$ is thus defined in
this limit from dynamical variables and is {\em not} prescribed from
the outside; $t$ controls the dynamics in this approximation.
\item The next order of the Born-Oppenheimer scheme yields quantum
  gravitational correction terms proportional to $G$ \cite{KS,OUP}.
  The presence of such terms may in principle lead to observable
  effects, for example, in the anisotropy spectrum of the cosmic
  microwave background radiation. These terms result from quantum
  dynamical features; quantum geometry may lead to additional
  corrections as seen in Section~3 on loop quantum cosmology.
\end{itemize}
The Born--Oppenheimer expansion scheme distinguishes a state of the
form (\ref{expiS}) from its complex conjugate. In fact, in a generic
situation where the total state is real, being for example a
superposition of (\ref{expiS}) with its complex conjugate, both
states will decohere from each other, that is, they will become
dynamically independent \cite{deco}. This is a type of symmetry
breaking, in analogy to the occurrence of parity violating states in
chiral molecules. It is through this mechanism that the i in the
Schr\"odinger equation emerges. Quite generally one can show how a
classical geometry emerges from quantum gravity in the sense of
decoherence \cite{deco}: irrelevant degrees of freedom (such as
density perturbations or small gravitational waves) interact with
the relevant ones (such as the scale factor or the relevant part of
the density perturbations), which leads to quantum entanglement.
Integrating out the irrelevant variables (which are contained in the
above multipoles $\{ x_n\}$) produces a density matrix for the
relevant variables, in which non-diagonal (interference) terms
become small \cite{decoqc1}. One can show that the Universe assumes
classical properties at the onset of inflation \cite{decoqc2}. The
quantum fluctuations out of which eventually the galaxies form
acquire classical properties through a similar mechanism \cite{KPS}.

Due to the linear
structure of quantum gravity, the total quantum state is a
superposition of many macroscopic branches even in the semiclassical
situation, each branch containing a corresponding version of the
observer (the various versions of the observer usually do not know of
each other due to decoherence). This is often referred to as the
``many-worlds (or Everett) interpretation of quantum theory'' \cite{Everett},
although only one {\em quantum} world (described by the full $\Psi$)
exists \cite{deco}.

\subsection{Arrow of time and structure formation}

Although most fundamental laws are invariant under time reversal,
there are several classes of phenomena in Nature that exhibit an
arrow of time \cite{Zeh}. It is generally expected that there is an
underlying master arrow of time behind these phenomena, and that
this master arrow can be found in cosmology. If there existed a
special initial condition of low entropy and if time proceeded in
terms of the scale factor $a$, statistical arguments could be
invoked to demonstrate that the entropy of the Universe will
increase with increasing size.

There are several subtle issues connected with this problem.
First, one does not yet know a general expression for the entropy
of the gravitational field; the only exception is the black-hole entropy,
which is given by the expression
\begin{equation}
\label{SBH}
S_{\rm BH}=\frac{k_{\rm B}A}{4G\hbar}=k_{\rm B}\frac{A}{4l_{\rm P}^2}
 \ ,
\end{equation}
where $A$ is the surface area of the event horizon, $l_{\rm P}$ is
again the Planck length and $k_{\rm B}$ denotes Boltzmann's
constant. According to this formula, the most likely state for our
Universe would result if all matter would assemble into a gigantic
black hole; this would maximize (\ref{SBH}), cf. [\refcite{vaas2}].
More generally, Roger Penrose has suggested\cite{PenroseEntropy} to
use the Weyl tensor as a measure of gravitational entropy, which
expresses the very special nature of the Big-Bang (small Weyl
tensor) and the generic nature of a Big-Crunch (large Weyl tensor).
Entropy would thus increase from Big-Bang to Big-Crunch. (See
[\refcite{Zeh}] for a detailed exposition and references.)

Second, since these boundary conditions apply in the very early
(or very late) Universe,
the problem has to be treated within quantum gravity.
But as we have seen, there is no external time in quantum gravity -- so
what does the notion ``arrow of time'' mean?

We shall address this issue in quantum geometrodynamics, but
  given that only the type of equations will be referred to the
situation should not be very different in loop quantum cosmology or
string cosmology.
An important observation is that the
Wheeler--DeWitt equation exhibits a
fundamental asymmetry with respect to the ``intrinsic time'' defined
by the sign of the kinetic term. Very schematically, one can write this
equation as
\begin{equation}
 {\mathcal{H}} \, \Psi =
\left(\frac{\partial^2}{\partial\alpha^2} + \sum_i \, \left[
-\frac{\partial^2}{\partial x_i^2}+\underbrace{V_i(\alpha,x_i)}_{\to 0\
{\rm for}\ \alpha
\rightarrow -\infty}\right]\right) \, \Psi = 0 \ ,
\end{equation}
where again $\alpha=\ln a$, and the $\{ x_i\}$ again denote
inhomogeneous degrees of freedom describing perturbations of the
Friedmann Universe (see above); $V_i(\alpha,x_i)$ are the potentials
of the inhomogeneities. The important property of the equation is
that the potential becomes small for $\alpha\to -\infty$ (where the
classical singularities would occur), but complicated for increasing
$\alpha$. In the general case (not restricting to small
inhomogeneities), this may be further motivated by the
BKL-conjecture according to which spatial gradients become small
near a spacelike singularity \cite{BKL}. The Wheeler--DeWitt
equation thus possesses an asymmetry with respect to ``intrinsic
time'' $\alpha$. One can in particular impose the simple boundary
condition
\begin{equation}
\Psi \quad \stackrel{\alpha \, \to \, -\infty}{\longrightarrow}\
\psi_0(\alpha)\prod_i \psi_i(x_i)\ ,
\end{equation}
which would mean that the degrees of freedom are initially {\em not}
entangled. Defining an entropy as the entanglement entropy between
relevant degrees of freedom (such as $\alpha$) and
irrelevant degrees of freedom (such as most of the $\{ x_i\}$), this
entropy vanishes initially but
increases with increasing $\alpha$ because entanglement increases
due to the presence of the potential. In the semiclassical limit where
$t$ is constructed from $\alpha$ (and other degrees of freedom),
cf. (\ref{semi}), entropy increases with increasing $t$. This, then, would
\emph{define} the direction of time and would be the origin of
the observed irreversibility in the world. The expansion of the
Universe would then be a tautology. Due to the increasing entanglement,
the Universe rapidly assumes classical properties for the
relevant degrees of freedom due to decoherence.
Decoherence is here calculated by integrating out the $\{ x_i\}$
in order to arrive at a reduced density matrix for $\alpha$
\cite{decoqc1,decoqc2}.

This process has interesting consequences for a classically
recollapsing Universe \cite{KZ,Zeh}. Since Big-Bang and Big-Crunch
correspond to the same region in configuration space
($\alpha\to-\infty$), an initial condition for $\alpha\to-\infty$
would encompass both regions. This would mean that the above initial
condition would always correlate increasing size of the Universe
with increasing entropy: the arrow of time would formally reverse at
the classical turning point; Big-Bang and Big-Crunch would be
identical regions in configuration space. As it turns out, however,
a reversal cannot be observed because the Universe would enter a
quantum phase \cite{KZ}. Further consequences concern black holes in
such a Universe because no horizon and no singularity would ever
form.

\subsection{Transition to the XXI$^{\mathrm{st}}$ century}

The main application to quantum cosmology in the last ten years is
motivated by loop quantum gravity and is described in the next
section. But work on the quantum geometrodynamical Wheeler--DeWitt
equation is also going on. This is at least in part due to the
reasons given at the beginning of my section above. But it is also
related to the fact that loop quantum cosmology, too, addresses a
constraint equation of the form ${\mathcal{H}}\Psi=0$, although it
is now a difference equation; for large-enough scale factors one
expects that this difference equation is approximately given by the
Wheeler--DeWitt equation.

Work on the standard Wheeler--DeWitt equation includes supersymmetric
quantum cosmology (see the remarks towards the end of our
contribution), path-integral methods (see, for example,
[\refcite{AJL05}]), or investigations of singularity
avoidance[\refcite{avoidance1,avoidance2,avoidance3}].
An overview of other recent developments can be found in [\refcite{Coule}].

%%%%%%%%%%%%%%%%%%%%%%%%%%%%%%%%%%%%%%%%%%%%%%%%%%%%%%%%%%%%%%%%%%%%%

\section{From the last decade(s) (Martin Bojowald)}
\label{sec3}

Several ones of the general issues in quantum cosmology are to be
faced by any approach, irrespective of its details and
technicalities. Among those are, starting at a rather fundamental
level and proceeding to more practical problems, (i) the
interpretation of the wave function of the Universe and of the
observable ingredients it contains, (ii) quantum dynamics of the
wave function formulated by a Wheeler--DeWitt-type equation, (iii)
the role of ``quantum geometry'' underlying the dynamical concepts
and the associated understanding of quantum space--time, and finally
(iv) the detailed prescription of feasible test procedures and
potential observational consequences of the whole framework.

Many of these questions, in accordance with their general nature, have
already been addressed in depth in Sec.~\ref{sec2}. But during the
last one or two decades, some of these issues have been approached
specifically within the realm of loop quantum gravity
\cite{Rov,ALRev,ThomasRev}, providing several new insights.
Especially the background independent notion of quantum geometry,
realized within this framework not just in reduced models of quantum
cosmology but in a general setting, has emphasized the importance of
point (iii) above.  But all these questions are interlinked, and thus
new ingredients have resulted from loop quantum gravity also for the
other issues.

By a change of perspective, loop quantum gravity has made significant
progress regarding one of the major issues that has so far stubbornly
remained out of reach for Wheeler--DeWitt quantizations: the rigorous
formulation of a {\em kinematical} quantum representation for general,
unrestricted geometries, going beyond exactly symmetric models or
perturbative multipole expansions around them.  Seen from the general
viewpoint of quantum field theory, the loop representation implements
background independence by introducing basic operators without
reference to a background space--time metric.  As always, basic
canonical fields should be smeared (that is, spatially integrated) for
well-defined quantum representations; otherwise delta-functions with
their infinities appear in the classical Poisson brackets.

For a background-independent quantization of gravity, the smearing
must be done in such a way that no metric other than the physical one
is used for integration measures. The only possibility known (so far)
has been provided by loop quantum gravity: \cite{LoopRep} use
holonomies $h_e(A)={\cal P}\exp(\int_e A^i_a\dot{e}^a\tau_i{\rm d} t)$
of the Ashtekar connection $A_a^i$ along spatial curves $e$ and fluxes
$F^{(f)}_S(E)=\int_SE^a_in_af^i{\rm d}^2y$ of the densitized triad
$E^a_i$ along spatial surfaces $S$ (with su(2)-generators $\tau_i$
proportional to the Pauli matrices and surface-supported smearing
functions $f^i$).  The fields $A^i_a$ and $E^a_i$ are canonically
conjugate to each other, giving rise to the holonomy-flux algebra
under taking Poisson brackets. A unique \cite{LOSTF} diffeomorphism
covariant quantum representation results, in which fluxes,
representing spatial geometry via the densitized triad, turn out to
have discrete spectra\cite{AreaVol}, and holonomy operators are not
continuous in the edge length. Unlike in the Wheeler--DeWitt
representation, connection (or curvature) components cannot be
represented directly.

These are the main lessons from loop quantum gravity that loop quantum
cosmology\cite{LivRev} attempts to incorporate in cosmological models.

\subsection{Difference equation}

The Wheeler--DeWitt equation (\ref{whdw1}) is obtained by quantizing
$a$ (or $\alpha=\ln a$) and its momentum in the well-known quantum
mechanical way. This procedure is not compatible with what we have
seen from full loop quantum gravity, where connections, and thus the
momenta of metric components, cannot be represented directly but
must rather refer to holonomies.

In an isotropic context, the connection reduces to
$A_a^i=c\delta_a^i$, conjugate to an isotropic densitized triad
$E^a_i=p\delta^a_i$. For spatially flat models, the isotropic
connection component is related to the scale factor by $c\propto
\dot{a}$, and it is canonically conjugate to $p$, determining the
metric via $a=\sqrt{|p|}$. Loop quantum cosmology, following full loop
quantum gravity, then provides a quantization only of isotropic
holonomies as functions of the isotropic connection component
$c\propto\dot{a}$, {\em not of $c$ directly}. Only functions of the
form $\exp(i\delta(a)\dot{a})$, as matrix elements of holonomies along
straight curves of length $\delta(a)$, can be turned into
operators.\footnote{One might expect the length parameter $\delta$ to
  be a constant, such as a regulator chosen once and for all. However,
  in a reduction of quantum gravity states, $\delta$ arises from the
  lengths of links in lattice-like structures, which are being refined
  as the dynamics of an expanding Universe unfolds. In minisuperspace
  models, this can be faithfully mimicked only by working with a
  phase-space dependent edge length, such as
  $\delta(a)$.\cite{InhomLattice}}

One can view non-linear holonomies as contributing to
higher-curvature terms, the leading order reproducing the Friedmann
dynamics. Adding suitable powers of $\delta(a)\dot{a}$, which are
small when the Hubble distance $a/\dot{a}$ is large compared to the
comoving edge length $a\delta(a)$, makes the Friedmann equation loop
quantizable. Once quantized, the exponentials of holonomies act as
shift operators on the spectrum $\{\mu\}$ of $\hat{p}$. As the
resulting dynamical equation for wave functions $\psi_{\mu}(\phi)$
depending on matter fields $\phi$ as well as triad eigenvalues $\mu$
one thus obtains a discrete evolution equation, a difference
equation of the form\cite{DiffRefs}
\begin{equation} \label{Diff}
C_+(\mu)\psi_{\mu+\delta(\mu)}(\phi)+C_0(\mu)\psi_{\mu}(\phi)+
C_-(\mu)\psi_{\mu-\delta(\mu)}(\phi)=
{\cal H}_{\phi}(\mu)\psi_{\mu}(\phi) \,.
\end{equation}
All coefficients, including the matter Hamiltonian ${\cal
  H}_{\phi}(\mu)$, can be computed explicitly for a specific choice of
the regularization and for the lattice-refining curve parameters
$\delta(a)$. Owing to these choices, the coefficients are subject to
quantization ambiguities. Several qualitative aspects of the
difference equation and its solutions are nevertheless robust.  When
the discreteness is not relevant, e.g.\ in low-curvature regimes in
which the wave function does not oscillate strongly, one can
Taylor-expand and reproduce the Wheeler--DeWitt equation as its
continuum limit \cite{SemiClass}.

At this stage, the usual wave function issues of quantum cosmology
arise.  But there are also new ones of mathematical nature,
related to an analysis of the resulting difference equations.
Especially in anisotropic models\cite{HomCosmo} quantizing the Bianchi
cosmologies, these equations can be rather more complicated than the
isotropic one shown here. Numerical techniques, especially for the
non-equidistant types of difference equations that arise for
complicated forms of $\delta(\mu)$ in (\ref{Diff}), are being
developed.\cite{RefinedNumeric} Another recent development is the
formulation of a path-integral picture of loop quantum cosmology, in
this case making contact with spin foam models as the covariant
version of loop quantum gravity.\cite{LQCSFM}

\subsection{Singularity resolution}

With its new mathematical structure, loop quantum cosmology also
makes possible applications regarding physical space-time effects. A
basic new phenomenon brought about by the difference-equation nature
of the dynamics of loop quantum cosmology concerns the singularity
problem. Classically, the initial-value problem underlying isotropic
cosmological models becomes ill-posed when certain quantities
diverge at the Big-Bang singularity. No unique extension beyond the
singularity can be found for a classical geometry, and so the
singularity appears as an insurmountable border to space-time and to
our knowledge of earliest stages of the Universe.

In quantum cosmology, a similar problem is to be addressed for the
wave function, which replaces space--time geometry as the
information carrier.  Sometimes, arguments for singularity
resolution can be put forward in the Wheeler--DeWitt context. But
generic statements are difficult to obtain, and there are models in
which this quantization does not resolve the singularity problem:
all states follow exactly the classical trajectories into the
singularity.\footnote{Specifically, a free massless scalar field in
a
  spatially flat isotropic space-time presents a solvable system free
  of quantum back-reaction. Quantum states follow exactly the
  classical trajectories and cannot avoid the singularity.}  Stronger
ingredients are needed.

In loop quantum cosmology, new qualitative features arise. First,
just kinematically we have an extension of minisuperspace across the
singularity: the space of isotropic geometries is given not by a
half-line $a>0$ with the singular $a=0$ at a boundary (or pushed to
infinity by $\alpha=\ln a$), but by the whole real axis of
densitized-triad components $p\in {\mathbb R}$. (Both signs are
allowed for $p$ since it is the component of a densitized triad,
which changes sign under a reversion of orientation.)  Physics might
still break down at $p=0$ (or $\mu=0$ at the quantum level), but
there is now a clear way of finding out what happens by analyzing
the difference equation. It turns out that the classical singularity
is resolved:\cite{Sing} any evolving wave function continues through
$\mu=0$. Starting from initial values for the wave function
somewhere in a well-understood regime, dynamics extends it uniquely
across the classical singularity.  Moreover, initial conditions for
solutions follow automatically; they are derived dynamically, not
imposed by hand.\cite{DynIn}

From the wave function, one may attempt to reconstruct the behavior
of space--time beyond the singularity. In general the geometry may
not be of the classical form we know. Here, one should use
observables rather than the wave function, but that analysis is not
required for a general statement of singularity resolution: Any
wave function is uniquely extended, and even if we lack complete
observables or an explicit physical inner product, we know that the
wave function, and thus quantum geometry, exists beyond the
classical singularity in a unique way.

In more specific models, fixing even the matter content, further
properties of the mechanisms of singularity resolution can be
elucidated. There are two different kinds of statements: (i) A
geometrical notion of {\em discrete internal time} $\mu$, as
embodied by the difference equation, implies an upper bound for
energy density; wave functions must have a minimum wave length to
find support on the discrete time lattice.  (ii) The gravitational
force then becomes repulsive at high densities, sometimes resulting
in a ``bounce''. Then, one obtains mean values for, say, the volume
in a dynamical state that do not collapse into the classical
singularity but are turned into re-expansion.  Showing that a
bounce\cite{BounceReview} in the strict sense is realized, based on
details of quantum geometries, is more difficult and obtained in
fewer models\cite{QuantumBigBang,APSCurved,NegCosNum}, than showing
the boundedness of energy densities.  Several main issues are to be
addressed for geometrical pictures of singularity resolution:
linking dynamical wave functions to observables, proper
normalizations of states by a physical inner product, properties of
sufficiently general states as they may be realized in generic
quantum regimes near a classical singularity, and the sensitivity to
perturbations.

The first bounce solutions at the physical Hilbert-space level of
loop quantum cosmology were obtained numerically in models with a
free, massless scalar as matter source.\cite{QuantumBigBang}
Resulting bounce pictures looked surprisingly tame, with hardly any
spreading or deformations of semiclassical wave packets throughout
the whole, supposedly violent Big-Bang phase. Initially, this seemed
exciting, for it suggested a comparatively simple analysis of
everything concerning the Big-Bang.  However, it soon became clear
that the models used, as well as the states considered, were rather
special: they are very close to exactly solvable, harmonic systems,
ones with only weak quantum back-reaction.

The underlying exactly solvable bounce model\cite{BouncePert} is
obtained for a spatially flat isotropic Universe. Such a model can
be formulated in a rather general way based on a power-law
assumption $\delta(a)=\delta_0a^{2x}$ for the lattice-refinement
behavior in (\ref{Diff}), parameterized by a real number $x$
(negative for refinement rather than coarsening to take place). If
we define the variables $V:=a^{2(1-x)}$ and $J:=a^3\exp(i\delta_0
a^{2x} \dot{a})$, which obey a linear Poisson algebra, the
difference equation of loop quantum cosmology with an energy density
$p_{\phi}^2/2a^3$ can be seen to equate the scalar momentum to
$p_{\phi}\propto |{\rm Im}J|$.  Now taking $\phi$ as internal time,
instead of the scale factor as often used in the Wheeler--DeWitt
context, $p_{\phi}$ is realized as a linear Hamiltonian. Linear
models are harmonic: like the harmonic oscillator they are free of
quantum back-reaction. Only a small number of parameters is required
to understand the evolution of expectation values in any state. With
the corrections from loop quantum cosmology, all solutions for
$\langle\hat{V}\rangle(\phi)$, computed in a state required only to
be semiclassical once, bounce.

As an exactly solvable one, this model is very special, as special as
the harmonic oscillator in quantum mechanics. Rather than exhibiting
general properties, such models are important as the basis of a
systematic perturbation theory\cite{EffAc}, the canonical
generalization of low-energy effective actions.  In quantum cosmology,
this gives rise to an effective Friedmann
equation\cite{QuantumBounce}
\begin{equation}\label{EffFried}
 \left(\frac{\dot{a}}{a}\right)^2 = \frac{8\pi G}{3}\left(\rho
 \left(1-\frac{\rho_Q}{\rho_{\rm crit}}\right)
 \pm\frac{1}{2}\sqrt{1-\frac{\rho_Q}{\rho_{\rm crit}}}
\eta (\rho-P)+ \frac{(\rho-P)^2}{\rho+P}\eta^2
\right)
\end{equation}
to describe the evolution of the scale factor $a$, now seen as the
expectation value obtained from a state in quantum cosmology. The
classical terms of the Friedmann equation are clearly recognizable,
showing that the classical limit is achievable in certain regimes:
$\eta$, parameterizing quantum correlations, must be small, and
$\rho_Q$, the energy density with quantum corrections from quantum
fluctuations, must be small compared to the critical density
$\rho_{\rm crit}=3/8\pi G(a\delta(a))^2$. In $\rho_{\rm crit}$, which
arises from holonomy corrections, the refinement function $\delta(a)$
enters.

We note that the derivation of Eq.~(\ref{EffFried}) relies on the
implementation of a physical inner product, or equivalently on using
appropriate reality conditions. In this way, one ensures that the
square root appearing in (\ref{EffFried}) is real.  An upper bound
$\rho_Q\leq \rho_{\rm crit}$ then follows irrespective of the state
realized at those densities.\cite{QuantumBounce} Secondly, if the
state is such that $\eta$ is very small when $\rho_Q\sim\rho_{\rm
  crit}$, then there is a bounce ($\dot{a}=0$) when energy densities
reach the critical value. In the very specific situation of a stiff
fluid with $\rho=P$, the case first analyzed
numerically\cite{QuantumBigBang}, a bounce at the critical density is
realized irrespective of the value of $\eta$. It remains unclear how
well the bounce result can be generalized to situations in which the
potential energy of matter dominates over the kinetic contribution.

\subsection{Observational contact?}

For reliable predictions, one must understand all effects and quantum
corrections, have some control on the quantum-to-classical transition,
and manage a systematic form of perturbation theory including
effective equations with quantum back-reaction and inhomogeneities. In
particular, it is not sufficient to base general expectations on what
has been seen to dominate only in a specific class of models. A
general and systematic analysis of the whole theory is required which,
needless to say, is only in its beginning stages for loop quantum
gravity.

Effective equations must be sufficiently general in order to deal with
proper dynamical states. There is no general reason to assume
near-Gaussian (or uncorrelated) states as they occur in expansions
around ordinary free field theory. A large class of underlying states
must be encompassed; and as seen in Wheeler--DeWitt models, results
can sensitively depend on the form of a state.\cite{WdWBounce}.

In order to implement inhomogeneities consistently, one must face the
anomaly issue, presenting strong consistency conditions from the usual
overdetermined set as given classically by Einstein's equations.
Simply modifying general relativistic equations is easily possible in
homogeneous models, but as soon as inhomogeneities are included,
perturbation equations must be delicately matched to the modified
background equations. If the corrected set remains consistent, the
system is anomaly-free.  For reliable conclusions, every quantum
correction considered in homogeneous models, for instance the
modification due to holonomies, must be shown to have a consistent
formulation with inhomogeneities. Since homogeneous equations can be
modified at will, as far as the anomaly problem is concerned,
producing specific bounce models or bounded densities is not that
difficult. What is non-trivial, and still lacking for several
quantum-geometry corrections, is a consistent embedding in a set of
anomaly-free equations for inhomogeneities.

The anomaly problem is related to the gauge problem: Fixing the
space--time gauge before quantization or before the derivation of
quantum corrections evades consistency issues and can be used to reduce
the over-determinedness of equations. But crucial effects can easily
be overlooked, especially since quantum gravity leads to corrections
of constraints which generate gauge transformations. Referring to the
gauge before one even knows the quantum corrected transformations can
easily be misleading. Several examples exist by now
in which perturbative inhomogeneities of certain forms can be
implemented
consistently\cite{ConstraintAlgebra,JR,LTBII}. The
resulting equations imply properties, such as non-conservation of
power on large scales or effective anisotropic stresses, that are
important for observational consequences but would not appear in this
form for gauge-fixed treatments. The anomaly problem makes any
analysis rather complicated, but it also provides us with a chance to
analyze quantum space--time on its smallest scales. Results then have
important implications for full quantum gravity.

%%%%%%%%%%%%%%%%%%%%%%%%%%%%%%%%%%%%%%%%%%%%%%%%%%%%%%%%%%%%%%%%%%%

\section{..., \emph{citoyens}! (All authors)}
\label{sec4}

The general set of queries and contributions (or concerns) from the
audience could be presented
in a somewhat fictionalized debate as follows:

{\em What exactly does it mean to quantize cosmology?} Most
investigations of quantum cosmology happen in a minisuperspace
framework, which starts with a classical truncation to finitely many
degrees of freedom, then quantizing them by techniques borrowed from
quantum mechanics (cf. Section~2).  In some cases, quantization has
been completed by arriving at a Hilbert-space of states with a
representation of a complete set of observables.  But those models
are based on global internal times from matter, such as a free
massless scalar or dust, ingredients which can hardly be considered
general. It remains unclear what mathematical structures are needed
for a general form of quantum cosmology.

{\em How can you at all trust results from those severely truncated
  quantum cosmological models?  } Indeed, quantum cosmology is a
truncation rather than an approximation, drastically cutting off
unwanted degrees of freedom instead of providing a harmonious
embedding of a simplified model within a fuller framework.  There is
currently no well-defined approximation scheme that would show under which
conditions terms ignored in full expressions could be considered
small.

The main difficulty is the lack of empirical tests by which to assess
the validity of approximations, following the example of theoretical
condensed matter physics which, too, employs severe truncations in
many cases. Instead, theoretical investigations have been
performed\cite{KR89}, more recently also using path
integrals \cite{AJL05}. Some support for minisuperspace
models comes from the BKL conjecture\cite{BKL}, according to which the
dynamics of space--time near a spacelike singularity is dominated by
its ultralocal behavior: time derivatives seem to dominate over
spatial ones. While fields can still have large spatial variations,
the evolution of the geometry at a point seems to depend only on the
geometry at that point. If this were true and held even in quantum
gravity, minisuperspace models would indeed capture the main dynamical
features at least in the approach to a spacelike singularity.

{\em What does it actually mean to ``resolve'' singularities?
  Shouldn't one consider curvature invariants and make sure that they
  all remain bounded?}  Curvature divergence is a
  feature of
most of the
solutions of general relativity, but it is
not a
property shown by singularity theorems; they use geodesic
incompleteness as the key criterion. Focusing on curvature
divergence for singularity resolution may thus be too restrictive or
even misleading. Alternatives are specific forms of boundary
conditions, or the quantum hyperbolicity condition of loop quantum
cosmology.

{\em
 How does one make sure that a state achieves
  semiclassicality at large volume?}  Semiclassical states play an
important role in developing quantum gravity, to check the correct
classical limit, and to find suitable regimes for low-energy physics.
In the Wheeler--DeWitt approach, the semiclassical limit is well
understood at a formal level, cf. Section~2.4.
Most methods to construct semiclassical states in loop quantum
gravity \cite{CohState} remain at the kinematical level (and thus do
{\em not} satisfy all constraints). Quantum cosmological models are often
simple enough so as to allow the construction of dynamical coherent
states and to see how semiclassicality takes place in a dynamical
context: how fast can an initial state spread or change shape?

There are, of course, all the conceptual issues discussed in
Section~2. When and how does decoherence act, that is, when can
superpositions of different semiclassical states be treated as
dynamically independent components? What is the relevance for the
arrow of time? It seems obvious that quantum cosmology is in severe
conflict with a Copenhagen-type of interpretation, which assumes the
presence of a classical world from the outside. Are there
alternatives to the Everett interpretation?

As also emphasized in Sections~2 and 3, quantum effects are not a
priori restricted to the Planck scale. They may even occur for large
Universes. Sizable quantum effects are responsible for the
singularity avoidance of scenarios containing a Big-Rip, a Big-
Brake, or other classically singular situations.  At small volume,
the dynamics of states may be much more violent than in
semiclassical regimes. A general discussion of intuitive
singularity-avoidance mechanisms, such as bounces, suffers
considerably from the lack of knowledge about dynamical
semiclassical states.

{\em Can quantum cosmology make realistic (that is, falsifiable)
  predictions?}  There are arguments that quantum cosmology can
predict inflation, combined with a reasonable spectrum of primordial
fluctuations; see [\refcite{BKKS}] and the references therein.  Also
bounce pictures in general terms lead to potential
signatures.\cite{CosmoWithoutInfl} Such statements are often related
to proposals for initial states, which can simultaneously be used to
address the singularity problem. One may also envisage quantum
gravitational corrections to standard cosmology scenarios, which
could exhibit themselves in the CMB anisotropy spectrum \cite{OUP}.
In this sense, quantum cosmology has been progressing in a
(self-)consistent manner facing the reality of astronomical
observations. However, a clear observation or a (currently feasibly
detectable) signature from a quantum Universe has not yet been
possible. Only the availability of clear-cut empirical tests would
confirm quantum cosmology as a viable approach to understand the
foundations of cosmology.

%%%%%%%%%%%%%%%%%%%%%%%%%%%%%%%%%%%%%%%%%%%%%%%%%%%%%%%%%%%%%%%%%%

\section{\emph{Nouvelle Vague}? (All authors)}
\label{sec5}

Some progress has already been made in understanding the central
conceptual issues such as the problem of time, the interpretation of
the wave function, or the connection of quantum cosmology with full
quantum gravity. As long as the full theory is not known, however,
these insights have to remain preliminary. If it turned out, for
example, that quantum gravity would be nonlinear
\cite{PenroseEntropy}, most (if not all) of these results (as well
as those of string theory) would become obsolete.

It is therefore at this stage quite difficult, if not impossible or
at least uncertain, to make either predictions or indicating
where quantum cosmology can be further developed in a significant
manner in the years (or decades) to come. In the following we provide
some possible lines, bearing in mind the risk of being proven
wrong by other advances or even observational evidence.

The ``(string) landscape'' issue has driven efforts from quantum
cosmology; see, for example, [\refcite{BouhmadiLopez:2006pf}] and
the references therein.  Assuming that it is a robust property of
string theory, one expects that the landscape picture requires
elements of statistical theory and quantum selection rules for
transitions. Could it be that quantum cosmology would merely provide
an ``average'' perspective and that a more ``field theory''-like
structure (e.g. a third quantization or inputs from statistical
physics) would benefit quantum cosmology and the manner in which the
(wave function of the) Universe is discussed? These are wide-open
issues.

We point out the interesting contribution of supersymmetric quantum
cosmology
\cite{Kiefer:2005qs,Moniz:1996pd,VargasMoniz:2003xq,Moniz:2000uh,Moniz:1997zz,D'Eath:2002ec,D'Eath:1996at}. 
It brings additional structure to the framework and still has
potential for fresh ideas, or even new problems with tentative
solutions, especially on how we probe the very structure of
space-time\cite{Kiefer:2005qs}. If supersymmetry is discovered at
particle accelerators, investigating the early Universe within any
quantum cosmology school may require the implementation of
supersymmetry.

In loop quantum cosmology, the link to the full theory has turned out
to be essential, even though so far it is rather weakly built. Details
of the equations of loop quantum cosmology can often be restricted by
knowing how ingredients may arise from the equations of full loop
quantum gravity. Effects are then important not just at ultra-high
densities but even in tame regimes, for instance, in the context of
the correct classical limit. Issues such as lattice refinement or
anomaly-free extensions to include inhomogeneities make crucial use of
properties of the fully theory --- and then provide important feedback
on the feasibility of full constructions.

Let us finally quote some questions for which we expect to obtain
(or think we already know) an answer from quantum cosmology:
\begin{itemize}
\item How does one have to impose boundary conditions?
\item Is the classical singularity really being avoided, and how so?
\item Will there be a genuine quantum phase in the future?
\item How does the appearance of our current classical Universe follow?
\item Can the arrow of time be understood from quantum cosmology?
\item How does the formation of structure proceed?
\item Can inflation itself be understood from quantum cosmology?
\item Can quantum cosmology
be justified
from
  full quantum gravity?
\item Which consequences can be drawn
for the interpretation of quantum theory in general and for
quantum information in particular?
\item Can quantum cosmology be experimentally tested?
\end{itemize}

We want to emphasize again that quantum cosmology will become an
established part of physics only if it can and will be
experimentally tested. We are optimistic but  we do not know when
this will happen. But let us finish by a quote from Erwin
Schr\"odinger \cite{Schroedinger} as a motivation for continuing to
ask questions about quantum cosmology: ``\ldots or else, one might
seriously worry that just where we forbid further questions there
could still be quite a bit worth knowing to ask
about.''\footnote{``\ldots sonst w\"are ernstlich zu bef\"urchten,
da\ss\ es dort, wo wir das Weiterfragen verbieten, wohl doch noch
einiges Wissenswerte zu fragen gibt.''}

%%%%%%%%%%%%%%%%%%%%%%%%%%%%%%%%%%%%%%%%%%%%%%%%%%%%%%%%%%%%%%%%%%

\vspace*{-0.3cm}\section*{Acknowledgements} \vspace*{-0.2cm} MB and
CK  acknowledge support by grants from the Foundational
Questions Institute (FQXi).  MB was supported by NSF grant
PHY-0748336.  PVM thanks PDCT/P/FIS/57547/2007 and CENTRA-IST.

%%%%%%%%%%%%%%%%%%%%%%%%%%%%%%%%%%%%%%%%%%%%%%%%%%%%%%%%%%%%%%%%%

%%%%%%%%%%%%%%%%%%%%%%%%%%%%%%%%%%%%%%%%%%%%%%%%%%%%%%%%%%%

\end{document}